# Développement et analyse multi outils d'un protocole MAC déterministe pour un réseau de capteurs sans fil

**Thierry Val — Adrien van den Bossche**

*LATTIS EA4155, Groupe SCSF, Université de Toulouse*
*IUT de Blagnac - UTM*
*BP60073, 1 place Georges Brassens, F-31703 Blagnac*
*{val, vandenbo}@iut-blagnac.fr*

RÉSUMÉ. *Nous présentons dans cet article une méthodologie multi outils de développement et d'analyse de protocole utilisée pour valider une nouvelle méthode d'accès. La technologie IEEE 802.15.4 / ZigBee sert de base protocolaire à la proposition d'une couche MAC déterministe offrant un haut niveau de QdS. Ce type de WPAN peut typiquement être utilisé pour des réseaux de capteurs sans fil à fortes contraintes temporelles. Afin de valider les protocoles proposés, trois outils complémentaires et adéquats sont utilisés : les Réseaux de Petri pour la validation formelle du séquencement des trames, un simulateur spécifique pour les aspects temporels et des métrologies sur un prototypage réel à base de composants ZigBee FREESCALE pour la caractérisation fine des couches physique et liaison.*

ABSTRACT. *In this article, we present a multi-tool method for the development and the analysis of a new medium access method. IEEE 802.15.4 / ZigBee technology has been used as a basis for this new determinist MAC layer which enables a high level of QoS. This WPAN can be typically used for wireless sensor networks which require strong temporal constraints. To validate the proposed protocol, three complementary and adequate tools are used: Petri Nets for the formal validation of the algorithm, a dedicated simulator for the temporal aspects, and some measures on a real prototype based on a couple of ZigBee FREESCALE components for the hardware characterization of layers #1 and #2.*

MOTS-CLÉS : *Réseau de capteur sans fil, IEEE 802.15.4, ZigBee, Méthode d'accès au médium, Qualité de Service, Prototypage, Validation formelle, RdP, Simulation, Analyse de performance, métrologie.*

## 1. Introduction

La conception de protocoles de communications est un domaine de recherche à part entière. De nombreuses méthodes et outils ont été développés et apportent des aides évidentes aux chercheurs et industriels. Suivant les compétences des acteurs et leurs origines, on utilise initialement de façon privilégiée plutôt des outils de spécification et de validation formelle, ou plutôt des outils de simulation de réseaux et de protocoles, voire du prototypage réel. Chaque technique apporte son lot d'avantages en fonction des besoins et du niveau d'abstraction utilisé. Plus les protocoles à développer et valider sont proches des couches basses, plus il semble naturel d'utiliser un niveau de granularité fin, proche du matériel et des caractéristiques intrinsèques des médiums. Ceci est d'autant plus vrai lorsque l'on parle d'ingénierie des protocoles pour des réseaux sans fil où les caractéristiques dynamiques des médiums immatériels sont prépondérantes.

Les auteurs de cet article ont été confronté à ce genre de choix d'outils et de méthodes, lors de la proposition et la validation d'une nouvelle méthode d'accès dédiée à un réseau sans fil de capteurs pour des applications à très fortes contraintes temporelles, imposant un déterminisme fort de cette couche MAC. Nous avons alors décidé d'utiliser de façon ciblée et complémentaire plusieurs techniques, adéquates, dans le développement du protocole. Cet article présente donc tout d'abord les caractéristiques du WPAN ZigBee/802.15.4 qui a servi de base à nos réflexions. Après avoir identifié ses lacunes face à notre cahier des charges, nous présentons notre proposition de nouvelle couche MAC déterministe. La plateforme de validation multi outils est ensuite détaillée en présentant l'utilisation des RdP pour l'aspect formel de la validation, la conception d'un simulateur dédié et les analyses de performances temporelles qui en découlent, et enfin, les métrologies liées à un prototypage réel sur composants ZigBee de Freescale/Motorola.

## 2. Présentation de la norme de communication IEEE 802.15.4/ZigBee

### 2.1. *Généralités*

Nos travaux se basent sur la technologie de communication sans fil courte portée IEEE 802.15.4/*ZigBee* (ZigBee Alliance, 2005) (Val *et al.,* 2008), qui propose une norme pour des communications bas débit pour des entités embarquées (réseaux de capteurs par exemple). Un émetteur/récepteur ZigBee est caractérisé par une portée de 10 à quelques centaines de mètres et un débit de 20 à 250 kbit/s. La norme prévoit l'utilisation de trois bandes de fréquence (868, 915 ou 2400 MHz). La spécification ZigBee propose une pile protocolaire propriétaire et légère, déclinable dans plusieurs versions. Elle s'appuie sur la norme IEEE 802.15.4 (IEEE, 2003) pour les couches *Physique* et *Liaison de données*. Elle propose ses propres couches supérieures (*Réseau*, etc.). ZigBee réalise de fortes économies d'énergie grâce à une optimisation des périodes de mise en veille du matériel (Freescale, 2005).

## 2.2. *Accès au médium radio*

La norme IEEE 802.15.4 prévoit deux modes complémentaires pour l'accès au médium radio : un mode avec contention, de type *Best Effort*, par utilisation du protocole CSMA/CA (*Carrier Sense Multiple Access with Collision Avoidance*) ; et un mode avec réservation de ressources par créneaux temporels garantis (GTS, *Guaranteed Time Slots*), où seul l'élément détenteur du GTS est autorisé à émettre.

Ce dernier mode permet d'entrevoir une approche de Qualité de Service garantie par un élément *coordinateur du réseau* ; cet élément est chargé de répartir les accès au médium des entités communicantes qui lui sont rattachées (on parle alors de topologie *en étoile*) dans une structure temporelle appelée *supertrame*. Pour distribuer les temps de parole, le coordinateur diffuse à la demande ou régulièrement des trames balises, ou *beacons*, permettant la synchronisation des nœuds communicants et la diffusion des informations relatives à la distribution des temps de parole. L'espace inter-balises ou supertrame comprend une période *active* où les éléments échangent les messages et une période *inactive*, optionnelle. La période active de la supertrame est divisée en 16 slots temporels de longueur égale ; le *beacon* du coordinateur occupe le début du premier slot, suivi par la période des accès avec contention, ou CAP (*Contention Access Period*), dans laquelle les accès au médium respectent le protocole CSMA/CA slotté. Si les feuilles en ont fait la demande en CSMA/CA auprès du coordinateur, les GTS occupent les derniers slots de la supertrame dans la CFP (*Contention Free Period*).

## 2.3. *Les faiblesses de la méthode d'accès face à notre problématique de recherche*

IEEE 802.15.4 propose donc un mécanisme de réservation de ressources permettant à quelques feuilles privilégiées de s'affranchir des collisions, principale source d'incertitudes temporelles au niveau *Liaison de données*. Nous souhaitons développer une couche MAC qui offrira un accès au médium entièrement garanti sans aucune collision potentielle. Compte tenu de ce cahier des charges arbitraire très contraignant, nous proposons de renforcer ce mécanisme par la mise en œuvre d'une couche MAC à forte QdS, comme cela peut être requis dans le cadre d'un réseau de capteurs ou, plus largement, dans le cadre de réseaux industriels sans fil (Lepage, 1991). Les améliorations proposées se justifient par plusieurs faiblesses identifiées comme telles compte tenu de nos objectifs.

– L'obtention d'une portion temporelle dédiée est conditionnée premièrement par la non saturation préalable du réseau. Sa capacité n'est pas infinie, mais la norme ne fournit aucune possibilité au coordinateur de *réserver à priori* une partie des ressources pour certains nœuds *critiques* d'un point de vue applicatif. Selon la norme IEEE 802.15.4, les premiers demandeurs sont les premiers servis, ce qui n'est pas une politique de répartition acceptable compte tenu de nos objectifs de déterminisme fort. De plus, le processus de demande, initié par l'élément désirant obtenir un GTS, génère un message envoyé au coordinateur dans la CAP, en mode *Best Effort*, donc sans aucune garantie de succès dans un temps borné.

– Si il est possible d'augmenter la longueur d'un GTS (de un à plusieurs slots) dans chaque supertrame, la norme ne prévoit pas en revanche une possibilité d'allocation moins fréquente, par exemple un GTS toutes les deux supertrames, de manière à conserver un accès dédié, mais moins fréquent ; les GTS reviennent périodiquement à la fréquence de la supertrame, fixée par le paramètre BO (*Beacon Order*), ce qui peut se révéler très consommateur de bande passante. Autrement dit, il n'est pas prévu, dans le mode sans contention, de pouvoir faire cohabiter des trafics présentant des physionomies différentes sans "gâcher" des slots.

– Si plusieurs coordinateurs cohabitent dans une même zone de portée, il y a risque de collision, même dans le cadre des GTS. En effet, la norme ne prévoit pas de mécanisme de communication entre coordinateurs ; il est alors possible que plusieurs coordinateurs voisins attribuent le même slot à l'une de leurs feuilles qui risquent alors d'entrer en collision les unes avec les autres.

Le chapitre 3 présente les principes et les améliorations que nous avons mis en place, pour fiabiliser cet accès au médium.

## 3. Proposition protocolaire d'une nouvelle couche MAC déterministe

### 3.1. *Eléments du système de communication et liens entre éléments*

Notre proposition prévoit trois catégories d'éléments communicants :

– un supercoordinateur (*PAN Coordinator*) unique. IEEE 802.15.4/ZigBee prévoient cette entité spéciale dont le rôle principal est d'attribuer des adresses. Dans notre proposition, cet élément se voit doté de nouvelles responsabilités pour la synchronisation et l'attribution des GTS. Notre solution consiste en une centralisation de toutes les demandes de réservation qui sont acheminées via les coordinateurs jusqu'au supercoordinateur qui dispose d'une connaissance exhaustive de tous les slots réservés. Cette entité centrale est à portée radio de tous les coordinateurs ; cette hypothèse est rendue peu restrictive par l'utilisation de modules IEEE 802.15.4 dotés d'amplificateurs comme le XBEEPRO (MaxStream, 2006) offrant une portée de l'ordre du kilomètre en champ libre (ce qui est largement suffisant pour grand nombre d'applications). De plus, cette centralisation permet de régler le problème du coordinateur caché. Chaque coordinateur se voit attribué un slot dédié à l'émission de sa balise, réglant ainsi les collisions de *beacons*. Nous appellerons ces « GTS dédiés aux *beacons* » : GBS (*Guaranteed Beacon Slot*),

– un coordinateur pour chaque étoile. Chaque coordinateur est à portée radio de ses entités terminales qu'il connaît à priori grâce à l'applicatif. Il peut aussi communiquer avec d'autres coordinateurs et avec le supercoordinateur (par un trafic garanti dans notre proposition). Le coordinateur relaie les demandes de GTS de ses feuilles vers le supercoordinateur,

– une ou plusieurs entités terminales, ou feuilles (capteurs / actionneurs) par étoile. Chaque feuille ne peut dialoguer qu'avec son coordinateur en CSMA/CA ou grâce à ses GTS et PDS attribuées par son coordinateur. Un PDS (*Previously Dedicated Slot*) est un GTS attribué au préalable pour assurer une entrée déterministe dans le réseau. (cf. 3.2.4)

La topologie compte donc deux types de communication : entre le supercoordinateur et ses coordinateurs, et entre chaque coordinateur et ses feuilles

**3.2. *Organisation de l'accès déterministe au médium***

3.2.1. *Notion de supertrame*

Dans notre proposition, seule la supertrame du supercoordinateur commence au slot 0 et se termine au slot 15. Les supertrames émises par les coordinateurs d'étoiles commencent à un slot *i* et se terminent à un slot *i* – 1 (modulo 16), i étant le slot du GBS. Nous appelons *superbeacon* le *beacon* émis par le supercoordinateur ; le *superbeacon* occupe chaque slot 0. Les *beacons* sont répartis entre deux *superbeacons* comme l'illustre la figure 1.

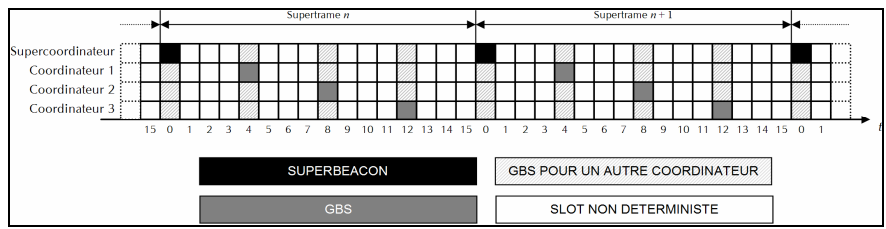

**Figure 1.** *Répartition des balises via les GBS dans la supertrame*

Ici, la supertrame du coordinateur 1 commence au slot 4 et se termine au slot 3 suivant. Pour que les *beacons* puissent être diffusés sans risquer d'entrer en collision avec un autre message, les slots qu'ils occupent leurs sont dédiés par le supercoordinateur qui choisit la valeur de *i* pour chaque coordinateur. Ainsi, le supercoordinateur peut répartir les émissions des *beacons* dans la supertrame, par exemple en la divisant en parts égales comme illustré par la figure 1. La frontière entre CAP et CFP est désormais supprimée ; les GTS et GBS peuvent être positionnés sur n'importe quel slot de la supertrame, en fonction des slots déjà attribués. Le choix de cet ordonnancement est laissé à l'application (regroupement pour favoriser de longues plages de veille ou, au contraire, dispersion pour favoriser une plus large distribution temporelle). En plus des slots réservés, les *beacons* doivent désormais annoncer explicitement les slots dédiés aux accès en CSMA/CA.

3.2.2. *Niveau de réservation 'n' variable*

Dans 802.15.4, les GTS alloués reviennent dans chaque supertrame. Dans notre proposition, chaque GTS peut être présent, au choix, dans toutes les supertrames, dans une supertrame sur deux, une sur quatre ou une sur huit, etc. On parlera de plusieurs niveaux de réservation notés *n*, avec, pour le cas de notre étude la distribution *n* = 0, 1, 2 ou 3. La période P des supertrames comprenant un GTS de niveau *n* est définie selon $P = 2^n$. Ce dernier point va permettre à des trafics différents de pouvoir cohabiter au sein d'une même étoile en respectant les contraintes de QdS fixées par l'application.

### 3.2.3. *Mécanisme de demande de réservation du médium*

Pour pouvoir accorder un GTS à l'un de ses nœuds, un coordinateur doit d'abord envoyer une requête au supercoordinateur et obtenir une réponse positive car seul cet élément a une vision exhaustive de la répartition des slots pour les $2^{n_{MAX}}$ supertrames à venir. Les dialogues déterministes entre le supercoordinateur et les coordinateurs prennent place dans les GBS, slots dédiés par le supercoordinateur pour chacun de ses coordinateurs. De plus, les GBS, s'ils sont simplement utilisés pour la diffusion des *beacons*, constituent une perte importante de bande passante. Nous proposons ainsi d'inclure la demande de GTS directement dans les *beacons* émis par les coordinateurs.

### 3.2.4. *Allocations au préalable : concept de PDS*

Nos objectifs en terme d'accès déterministe au médium impliquent que toutes les opérations réalisées sur le réseau puissent être effectuées de manière déterministe. Notre proposition prévoit donc une possibilité de réservation de slots *au préalable* (PDS), c'est-à-dire avant même que certains éléments demandent leur attachement au réseau. Cette fonctionnalité est indispensable pour garantir les propriétés déterministes de notre réseau pendant toutes les phases de son fonctionnement, y compris au moment de sa création. Elle est rendue possible car le réseau considéré a une taille limitée (environ 30 nœuds par exemple). Toutes les entités potentielles peuvent être connues avant la phase de création. Le supercoordinateur peut, dès sa mise en fonction, réserver des slots pour certaines entités comme les coordinateurs ou les capteurs critiques d'un point de vue applicatif. Ces PDS peuvent également permettre un changement de coordinateur sans rupture du service d'accès au médium garanti, par exemple dans le cas d'un nœud mobile. Il est important de noter que tant que l'entité détentrice du PDS ne s'annonce pas, les PDS constituent une ressource perdue pour le système de communication. Cependant, comme les GTS et les GBS, les PDS sont attribués avec un niveau de réservation *n* en fonction de la criticité de l'élément concerné, permettant ainsi de jouer sur le rapport *période du PDS / ressources perdues*. Une étude théorique analysant les ressources perdues par un PDS inutilisé est détaillée dans (van den Bossche A., 2007).

### 3.2.5. *Accès simultanés : concept de SGTS (Simultaneous GTS)*

Dans certaines conditions de propagation, nous proposons également des GTS simultanés attribués dans deux étoiles suffisamment éloignées pour ne pas se perturber (van den Bossche A., 2007). Le protocole est le suivant : supposons un réseau constitué de deux cellules (1, 2) comportant chacune un coordinateur (C1, C2) et une feuille (F1, F2). Les deux feuilles dialoguant sans collision dans un GTS, le coordinateur de l'autre cellule peut alors tenter de capter le signal émis et évaluer la puissance avec laquelle il reçoit la feuille de l'autre cellule. Soit PF1 la puissance reçue par C1 de sa feuille F1 et PF2 la puissance reçue de F2 ; si PF1 est supérieure à PF2 + marge, alors C1 ne sera alors pas perturbé par F2 si F1 et F2 émettent en même temps. C1 peut alors proposer un SGTS (F1, F2) à C2 qui devra procéder de la même manière. Si les deux coordinateurs arrivent à la même conclusion, le supercoordinateur peut regrouper les GTS de F1 et F2 en un même SGTS, permettant une utilisation optimisée de la bande passante totale du réseau. Le prototypage et la métrologie effectuée sur les conditions de propagation et de réception, présentés en 4.3 nous ont permis de valider ce concept et d'en déterminer les conditions optimales (marge > 10 dB).

**4. Plateforme multi outils de développement et de validation protocolaire**

Les nouveaux concepts présentés dans cette couche MAC originale doivent être validés. Plusieurs niveaux sont à traités : des aspects protocolaires généraux pour détecter des blocages ou des boucles dans le séquencement protocolaires des trames, des analyses de performances du point de vue temporel, mais aussi des mesures de propagation et de réception radio. Il nous semblait assez illusoire de pouvoir utiliser uniquement un seul outil. Nous avons au contraire décidé d'employer pour chaque problématique le meilleur outil, offrant la granularité suffisante et le niveau de validation de plus grand, en tentant à chaque fois de prendre en compte également la simplicité et rapidité de mise en œuvre. La figure 2 présente l'utilisation conjointe de trois outils et méthodes permettant de valider au moins à chaque fois une problématique. Avant tout travail affiné sur les analyses de performances et sur la caractérisation de la couche physique, il nous a semblé indispensable de valider formellement le séquencement protocolaire de ce nouveau protocole MAC. C'est par une modélisation par différents Réseaux de Petri (RdP), associée à des méthodes analytiques et l'utilisation de l'outil TINA (Berthomieu, 2004) développé au LAAS par le groupe OLC (Laas, 2008), que nous avons validé les différentes phases protocolaire, en prouvant le caractère borné, vivant et réinitialisable des différents RdP élaborés. Nous présenterons à titre d'exemple en 4.1. le RdP ayant permis de valider formellement l'association déterministe dans le réseau sans fil.

Il existe depuis peu plusieurs outils de simulation (OPNET, NS2…) disposant de modèles associés au réseau ZigBee. NS2 inclut des modèles de 802.15.4 (Zheng J, 2006), (Samsung, 2007) depuis la version 2.27. Ces modèles, comme ceux d'OPNET, sont généralement des contributions de recherche qui sont maintenant incluses et distribuées avec les simulateurs. Ils ont été étudiés pour traiter un ou plusieurs points particuliers mais ne modélisent pas toute la norme 802.15.4 / ZigBee (et encore moins notre couche MAC). Dans notre cas, nous souhaitions disposer d'un outil simple de simulation afin d'évaluer les aspects temporels de nos contributions spécifiques sur la couche MAC, avec l'idée de se servir des algorithmes utilisés en simulation pour faciliter le prototypage et l'implémentation sur microcontrôleur. Il était alors bien plus rapide et efficace de concevoir un simulateur spécifique pour cela, l'idée finale étant de l'associer aux autres outils (RdP et prototype), bien plus performants et fiables pour valider les autres aspects. Nous présenterons en 4.2. les résultats de simulation obtenus à partir de notre simulateur pour caractériser la latence pour l'association déterministe par PDS. Ce simulateur nous a aussi permis d'évaluer les performances de l'accès en CSMA/CA dans la CAP, ce point là ne sera pas détaillé ici, présentant moins d'originalité dans notre contribution.

|  | Méthodes | | |
|---|---|---|---|
|  | Validation formelle | Simulation | Prototypage |
| Séquencement de l'algorithme (exemple de l'association déterministe) | ● |  |  |
| Association déterministe par PDS et latence constatée pour les coordinateurs |  | ● |  |
| Association déterministe par PDS et latence constatée pour les nœuds |  | ● |  |
| Performances de l'existant avec l'accès en CSMA/CA dans la CAP |  | ● |  |
| Evaluation de la sensibilité du récepteur |  |  | ● |
| Evaluation du temps de traitement d'un paquet par la pile SMAC |  |  | ● |
| Synchronisation des nœuds d'une même étoile – gigue quantifiée |  |  | ● |
| Synchronisation des nœuds du réseau par cascade de beacons – gigue quantifiée |  |  | ● |
| Capacité maximale d'un slot et évaluation du débit garanti par la méthode d'accès |  |  | ● |
| Caractérisation de l'effet de capture et validation des SGTS |  |  | ● |

**Figure 2.** *Validation multi outils et méthodes*

Les aspects protocolaires proches de la radio sont difficilement modélisables de façon parfaite avec des outils de validation formelle et des simulateurs de réseaux. Les aspects temporels liés aux vitesses d'exécution des microcontrôleurs sont souvent omis dans les simulateurs réseaux. La caractérisation des propagations radio et des phénomènes de captures, prépondérants pour une couche MAC, est également assez délicate à obtenir avec des outils de haut niveau. Notre idée a été, au contraire, de valider ces concepts matériels par un prototypage réel et une étude du médium radio. A titre d'exemple, nous présentons en 4.3 l'étude qui nous a permis de mettre en évidence l'effet de capture et d'en déterminer les paramètres de faisabilité des SGTS. D'autres problématiques ont été analysées et validées grâce au prototype : la sensibilité du récepteur radio, l'évaluation du temps de traitement d'un paquet de niveau SMAC (*Simple MAC*, interface Freescale au plus bas de la couche MAC), la qualité de la synchronisation des nœuds via les *beacons* (intra et extra étoile), et l'évaluation des débits utiles offerts par la couche MAC, en tenant compte des caractéristiques de la couche physique (taux d'erreur bit, temps de propagation, charge processeur, temps de sérialisation pour l'échange des SDU entre les couches 1 et 2…).

### 4.1. *Validation formelle par RdP : cas de l'association déterministe*

Lorsqu'un capteur (fils) désire se connecter au réseau, il doit s'associer à son coordinateur (père) au sein de son étoile. Ceci va lui permettre d'émettre des données en CSMA/CA et dans ses GTS obtenus. La présence, dans la supertrame, d'un PDS pour ce capteur, permet une association déterministe sans aucun risque de collision. Le RdP, présenté en figure 3, permet de modéliser ce fonctionnement.

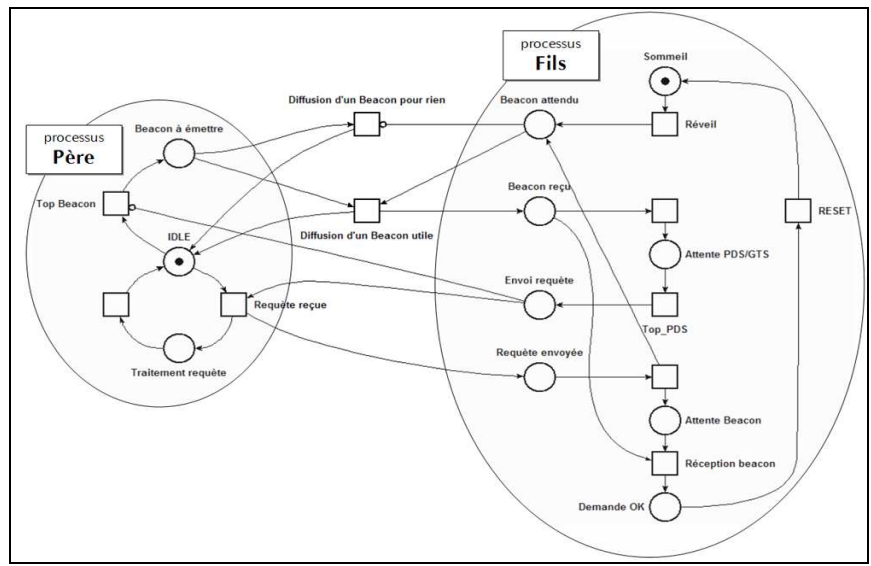

**Figure 3.** *RdP de l'association déterministe*

Le fils doit attendre la synchronisation émise de son père (*beacon* utile), puis son PDS, qu'il peut utiliser pour émettre des données, ou d'autres requêtes de *beacons*. Le principe est quasiment le même entre un coordinateur et le supercoordinateur. Les différents RdP, comme celui-ci présenté en exemple, ont été validé par des méthodes analytiques à l'aide de l'outil TINA. Les 3 aspects (borné, vivant et réinitialisable) ont été prouvés. Ces RdP et leur validation, ont permis de « défricher » les algorithmes de base de la méthode d'accès.

### 4.2. *Simulation : évaluation de la latence pour l'association par PDS*

Les différents algorithmes du protocole MAC, après avoir été validés formellement, ont servi de base à la conception d'un simulateur, écrit en langage C, puis ont été adaptés pour être prototypés. Ce simulateur s'est focalisé en particulier sur l'évaluation des latences du protocole dans la zone CFP de la supertrame. La figure 4 présente un des résultats obtenu par simulation. Nous pouvons identifier les différentes fenêtres de temps pour l'association au réseau d'un coordinateur, pour plusieurs valeurs de $n_{PDS}$ ($0 \leq n_{PDS} \leq 3$). Afin de se focaliser sur chaque problème un par un, ici, le médium radio est supposé parfait, c'est-à-dire sans erreur de transmission. L'axe des abscisses représente le temps nécessaire à l'association déterministe d'un coordinateur. Ce temps est découpé en intervalles car le système est temporellement discret : compte tenu du découpage en slots temporels, une demande d'association ne peut se faire qu'à certains instants. L'axe des ordonnées représente la proportion d'associations réalisées sur chaque tranche temporelle. Bien entendu, plus $n_{PDS}$ est grand (c'est-à-dire plus la fréquence des slots réservés au préalable est faible), plus la plage temporelle sur laquelle les associations sont réparties est large.

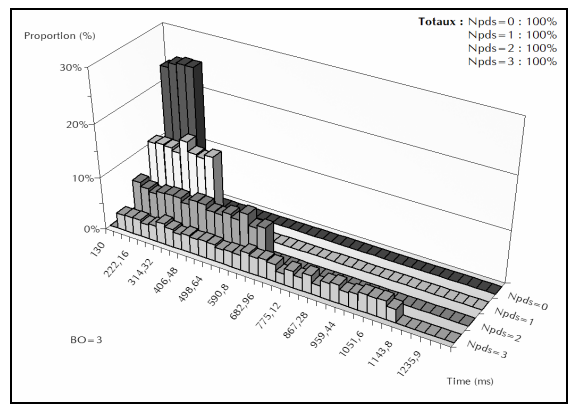

**Figure 4.** *Représentation des fenêtres temporelles pour l'association déterministe d'un coordinateur (pour BO = 3)*

Nous pouvons constater que la méthode d'association au réseau est déterministe puisque 100% des demandes obtiennent une réponse en un temps qui a une borne minimale et maximale. Il est possible pour un coordinateur, à partir du moment où il entre sur le réseau, d'être certain qu'il obtiendra une réponse sur son association dans un temps compris dans une fenêtre temporelle $FT_{COORD}$ telle que $17*T_{SLOT}*2^{BO} \leq FT_{COORD} \leq 16*T_{SLOT}*2^{BO}(2^{NPDS}+1)$, ce qui est impossible à borner en CSMA/CA.

**4.3.** *Prototypage et métrologie : validation de la faisabilité des SGTS*

C'est par prototypage et mesures radio que nous avons validé les concepts liés au médium radio et aux caractéristiques matérielles, en particulier le concept des SGTS. Nous avons conçu et programmé au niveau MAC des cartes à base de composants FREESCALE (Freescale, 2005). Ces modules sont composés d'un circuit émetteur/récepteur 2,4 GHz spécifique (MC13192) et d'un microcontrôleur 8 bits (MC9S08GT60). A l'inverse de cartes WiFi ou Bluetooth du commerce, il est possible, avec ces composants, d'attaquer la programmation au niveau le plus bas possible, juste au dessus de la couche radio. Nous avons ainsi pu reprogrammer complètement la couche MAC et des applications de tests et de mesures. Un des points les plus sensible à valider est le concept des SGTS. Pour cela, un prototype de métrologie de réception radio a été conçu autour de 5 nœuds : 1 supercoordinateur SC, 2 coordinateurs C1/C2, 2 feuilles N1/N2. Ni est associé à Ci. Tous les nœuds sont à portée radio, et les mesures ont été effectuées en chambre anéchoïque afin d'éliminer les bruits radio et limiter les réflexions. La figure 5 représente un résultat de mesure primordial pour prouver les conditions de faisabilité des SGTS.

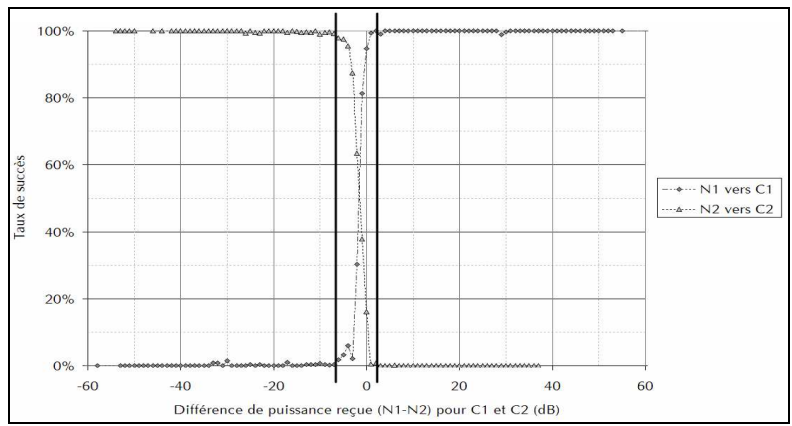

**Figure 5.** *Taux de trames reçues correctement en fonction du rapport de puissances reçues*

Le SC synchronise ses Ci par des *beacons* périodiques. Chaque Ci propage la synchronisation à sa feuille Ni. Chaque Ni émet alternativement une trame de données à son Ci (ceci permet d'évaluer la puissance de réception individuellement), puis les deux feuilles Ni émettent en même temps (SGTS) une trame. Ce processus est répété périodiquement en faisant varier les puissances d'émission des feuilles. La figure 5 représente le taux de trames reçues avec succès dans le SGTS en fonction de la différence de puissance reçue (N1-N2 pour C1, ou N2-N1 pour C2). Nous pouvons identifier une marge $\approx$ 10 db quasiment centrée sur 0 - avec un léger décalage du à une imperfection de synchronisation (van den Bossche, 2007) - en dehors de laquelle les SGTS sont efficaces puisque les trames de données sont reçues à pratiquement 100%. Ceci met en lumière le phénomène de capture où un récepteur se "cale" sur le premier signal reçu, même si un autre émetteur reçu légèrement plus fort débute son émission quelques instants après.

## 5. Conclusion

Le chapitre 4 de ce papier ne présente que quelques uns des nombreux résultats obtenus par les 3 méthodes et outils complémentaires pour valider cette nouvelle couche MAC dédiée aux réseaux de capteur sans fil. L'ensemble des résultats est disponible sur (van den Bossche, 2007). L'utilisation conjointe des outils de modélisation et validation par RdP, d'un simulateur de réseau dédié, et d'un prototypage réel associé à des métrologies, a montré son efficacité dans ce processus d'ingénierie de protocole et de réseau sans fil. Chaque outil apporte son avantage et son niveau de granularité adéquat. Les algorithmes protocolaires ont été successivement validés formellement grâce aux RdP, évalués temporellement par simulation, puis testés en environnement réel sur la maquette. Ce prototypage final a été grandement simplifié grâce aux 2 étapes précédentes qui ont permis de corriger toutes les erreurs de conception et de codage que l'on est malheureusement amené à faire dans la conception d'un protocole de niveau MAC aussi tributaire des contraintes temporelles. Il a été bien plus simple et efficace de valider sur la maquette, les problèmes radio et ceux liés aux charges processeur implémentant la couche 1 et 2, plutôt que de se lancer dans une modélisation complexe de la couche 2.4 GHz DSSS de 802.15.4, et du matériel informatique industriel embarqué dans les nœuds ZigBee Freescale. Enfin, même si les opérations de validation ont été établies dans l'ordre présenté dans cet article (RdP, puis simulation puis prototypage), nous envisageons cependant un retour arrière qui nous permettrait d'intégrer maintenant des résultats de métrologie radio et temporelles dans le simulateur, voire même dans les modèles RdP.

**Références**